\documentclass[aps,twocolumn,preprintnumbers,amsmath,amssymb,floatfix,groupedaddress,nofootinbib]{revtex4}

\usepackage{enumerate}
\usepackage{graphicx}
\usepackage{color}

\usepackage[applemac]{inputenc} 
\usepackage{fontenc}  
\usepackage{amsmath}

\newcommand{\beq}{\begin{equation}}
\newcommand{\eeq}{\end{equation}}
\newcommand{\bea}{\begin{eqnarray}}
\newcommand{\eea}{\end{eqnarray}}
\newcommand{\ba}{\begin{array}}
\newcommand{\ea}{\end{array}}

\newcommand{\bef}{\begin{figure}}
\newcommand{\eef}{\end{figure}}

\begin{document}

\title{Deriving Born's rule from an Inference to the Best Explanation.}

\author{Alexia Auff\`eves$^{(1)}$ and Philippe Grangier$^{(2)}$}

\affiliation{ 
(1): Institut N\' eel$,\;$BP 166$,\;$25 rue des Martyrs, F38042 Grenoble Cedex 9, France. \\
(2): Laboratoire Charles Fabry, IOGS, CNRS, Universit\'e Paris~Saclay, F91127 Palaiseau, France.}

\begin{abstract}
In  previous articles we presented a simple set of axioms named ``Contexts, Systems and Modalities" (CSM), where the structure of quantum mechanics appears as a result of the interplay between the quantized number of modalities accessible to a quantum system, and the continuum of  contexts that are required to define these modalities.  In the present article we discuss further how to obtain (or rather infer) Born’s rule within this framework. Our approach is compared with other former and recent derivations, and its strong links with Gleason's theorem are particularly emphasized. 
\end{abstract}

\maketitle

\section{Introduction.} 
\vskip -2mm

Many recent articles \cite{RSA} claim to provide new derivations of Born's rule, that is clearly a major theoretical basis of quantum mechanics (QM). Others claim that deriving Born's rule is a nonsense, and that it must essentially be postulated \cite{blog}. In this article we take a medium position, that is: Born's rule cannot be logically proven, but it does not have either to be postulated. 
Actually,  it can be inferred from some simple physical requirements or postulates, based on established (quantum) empirical evidence \cite{lipton}. After the realization of loophole free Bell tests \cite{AA} and in the era of quantum technologies \cite{N&C}, these postulates can be simply formulated \cite{csm1,CSM_Bell,CSM_Born1,CSM_random,CSM_Born2,csm4a}. They are presented here in a synthetic form, and various consequences are obtained and discussed.

\section{Definitions and postulates. } 
\vskip -2mm

{\bf Definition 1 : } {\it We consider a quantum system S and a specified ensemble of measurement devices interacting with it; this ensemble is called a {\bf context}\footnote{The word ``context" includes the actual  settings of the device, e.g. measurement of $S_z$ rather than $S_x$: the context must be factual, not contrafactual. On the other hand all devices able to measure $S_z$ are equivalent as a context, in a (Bohrian) sense that they all define the same conditions for predicting the future behaviour of the system.}. 
The best physically allowed measurement process provides a set of numbers, corresponding to the values of a well-defined and complete set of jointly measurable quantities; these values will be found again with certainty, as long as the system and context are kept the same\footnote{We omit the free evolution of the system; if it is present, the result of a new measurement can still be predicted with certainty, but in another context that can be deduced from the free evolution. {\it Mutatis mutandis}, this is equivalent to full repeatability.}. The physical situation occurring after such an ideal and repeatable measurement process is called a {\bf modality}.}

As an example using QM notations (not required yet),  for $K$ particles with spin 1/2,  the set of observables $\{ S_z^{(i)}, i = 1...K \}$ constitutes a context, and the observation of a given set of results $\{ m^{(i)},  i = 1...K \}$, where $m^{(i)} $ equals either $+\hbar/2$ or $-\hbar/2$,  constitutes a modality.  
The modalities are not defined in the same way as the usual ``quantum states of the system", since they are explicitly attached to both the context and the system.

From the above definition, justified by empirical evidence, one measurement provides only one modality. Therefore in any given context the various possible modalities are {\bf mutually exclusive}, meaning that if one result is true, or verified, all other ones are not true, or not verified. We have then the 

\vskip 2mm

\noindent {\bf Basic postulate (contextual quantization)} : {\it The number $N$ of mutually exclusive modalities for a given quantum system is the same in any relevant context. }  
 
\noindent  In the above example one has $N = 2^K$.  

\vskip 2mm

\noindent {\bf Definition 2 (incompatible modalities) :} {\it Modalities observed in different contexts are generally not mutually exclusive, they are said to be  incompatible.}   

\noindent Incompatible means that if a result is true, or verified, one cannot tell whether the other one is true or not.
\vskip 2mm

\noindent  {\bf Definition 3 (extravalence\footnote{In ref. \cite{CSM_Born1} extravalent modalities in different contexts are considered to be the same modality, transferred from a context to another. This is however not satisfactory, since a modality belongs to a specific context (and system). The notions of extracontextualy and extravalence are therefore more suitable to distinguish modalities and ``state vectors", as explained in ref. \cite{CSM_Born2}.}) :  }{\it When S interacts in succession with different contexts, certainty and repeatability can be transferred between their modalities. This is called {\bf extracontextuality}, and  defines an equivalence class between modalities, called {\bf extravalence\footnote{Note that extravalent modalities appear only if $N \geq 3$, this has an obvious geometrical interpretation in relation with Gleason's theorem (see below).}.}}

\noindent The equivalence relation is obvious, for more details and examples of extravalence classes  see  \cite{CSM_Born2}. 

The intuitive idea behind these definitions and postulate is that making more measurements in quantum mechanics (by changing the context) cannot  provide ``more details" about the system, because this would increase the number of mutually exclusive modalities, contradicting the basic postulate. One might conclude that changing context randomizes all results, but this is not true: some modalities may be related with certainty between different contexts, this is why extravalence is an essential feature of the construction. 

\section{Theorems.} 

{\bf Theorem 1 : } {\it Given an initial modality and context, obtaining another modality in another context must (in general) follow a probabilistic law.}
\vskip 1mm

First, let us emphasize that modalities in different contexts are always considered different, even if they are extravalent, so some care is required when counting modalities. 
Let us start from  an initial modality for a system in context $C_u$, and perform a measurement in another context $C_v$.  Several situations can be considered: 

(i)  From the basic postulate there are $N$ mutually exclusive modalities in each context, and one of them is realized when doing a measurement. Therefore the situation where all modalities in context $C_v$ would have a probability $p=0$ to occur is excluded by construction.

(ii)  If one modality in the new context $C_v$ is obtained with certainty, this means that $C_v$ contains a modality extravalent with the initial one; then $p = 1$ for this modality, and $p = 0$ for all other (mutually exclusive) ones. 
If the situation is the same for all modalities in $C_u$, then they are all extravalent with a modality in $C_v$, and the modalities in the new context can be seen as a rearrangement (permutation) of the initial ones. So let's try again with another context $C_w$; if the situation is the same again in all other contexts, it means that there are only $N$ classes of extravalent modalities, going through all contexts. This means that the context is unique up to a rearrangement (permutation) of the modalities; therefore there are no incompatible modalities, and the situation is essentially classical. 

(iii) Since case (i) is excluded, and case (ii) is classical (there are no incompatible modalities), the general case (where incompatible modalities do exist)  is that obtaining a modality in the new context is probabilistic $( 0 < p < 1)$, hence the theorem is demonstrated.  $\square$
\vskip 1mm
The core of this proof is that measuring in a new context cannot be a ``refinement''  of the previous measurement, because this would extend the number $N$ of mutually exclusive modalities. To see that more explicitly,  let us consider an initial modality $u_0$ in $C_u$, connected to at least two modalities $v_1$ or $v_2$, according to (iii) above. 
Now let us measure again in $C_u$: if $u_0$ is found again with certainty, then there would be two mutually exclusive situations, 
$u_0 \rightarrow v_1 \rightarrow u_0$ and $u_0 \rightarrow v_2 \rightarrow u_0$. This would give at least $(N+1)$ mutually exclusive modalities, in contradiction with the quantization postulate. 

Therefore the randomness is not only from $C_u$ to $C_v$, but also back from $C_v$ to $C_u$ \cite{csm4a}.  This makes clear that probabilities do follow from the fixed value of $N$, i.e. from the maximum number of mutually exclusive modalities for a given system, imposed by the basic postulate. 
\vskip 1mm

{\bf Theorem 2 : } {\it Given an initial modality and context, the probability to get another modality in another context keeps the same value as long as  the initial and final modalities belong to the same respective extravalence class, independently of the embedding contexts.}
\vskip 1mm

Let us start again from  an initial modality $u_i$ and context $C_u$, and follow the same steps as in Theorem 1 when performing a measurement in another context $C_v$.   
\vskip 1mm 

(i)  The situation where no modality can be obtained in the new context  $(p = 0)$ is  excluded as said above. 

(ii)  The situation where obtaining one modality in the new context is certain $(p = 1)$ means that the new context contains a modality extravalent with the initial one. Then $p=1$ corresponds to modalities in the same extravalence class, this is the definition of extravalence.

(iii) In the general case one gets another modality $v_j$ with a probability $0 < p < 1$. Given this new modality $v_j$, changing again the context to another one $C_w$ containing a modality $w_k$ extravalent to $v_j$ will yield $w_k$ with certainty. In that case the probability for going from $u_i$ to $v_j$ will be the same as the one for going from $u_i$ to $w_k$ (Fig. 1).  Moreover, if one starts from a modality $x_l$ extravalent to $u_i$, and one goes to $u_i$ then to $v_j$, the probability for going from $x_l$ to $v_j$ will be the same as the one for going from $u_i$ to $v_j$.  

Therefore the probability to get another modality in another context only depends on the extravalence classes of the initial and final modalities, and the theorem is demonstrated\footnote{In order to make sense of Theorem 2, it is essential to distinguish between modalities and vectors in an Hilbert space, that will correspond to extravalence classes of modalities (see below). This issue is also essential for understanding Gleason's hypotheses.}.  $\square$

\begin{figure}[h]
\includegraphics[width = 2.8cm]{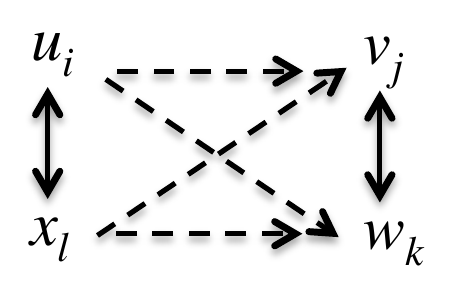}
\caption{
The four modalities $\{u_i, \, v_j, \, x_l, \, w_k \}$ belong to four different contexts,  and $u_i$  (resp. $v_j$) is extravalent with $x_l$ (resp. $w_k$). Then all probabilities represented by dashed lines are equal according to Theorem 2. }
\end{figure}

This theorem shows that the probability to get a new modality starting from an initial one is  linked neither to the context, nor to the modalities themselves, but to their extravalence class.  In some approaches this property is called ``non-contextual assignment of probabilities'', and this is a very fundamental feature of quantum mechanics, which appears here as a theorem. It also suggests the major next step, i.e. that the probability law should be obtained by attributing a mathematical object to an extravalence class, in such a way that all the above requirements are fulfilled.  As a general feature of such an inductive or inference reasoning \cite{lipton}, it cannot be shown that the proposed solution is unique (ie, necessary), but it can be shown that it fulfills all the requirements (ie, that it is sufficient). 
\vskip 2mm

{\bf Theorem  3 : } {\it  Let us associate a $N \times N$ rank-1 projector $P_i$ to each extravalence class of modalities, and a set of $N$ mutually orthogonal projectors to each context. Then the probability law $f(P_i)$ built from these projectors obeys Born's rule, and different sets of mutually orthogonal projectors are related by (complex) unitary matrices. }
\vskip 1mm 

Since the $N \times N$ projectors are associated to extravalence classes of modalities, the probabilities are a function $f(P_i)$ of these projectors, in agreement with  Theorem~2. Since a context (set of mutually exclusive modalities) is associated to a set of  $N$ mutually orthogonal projectors, the probabilities for this set of projectors sum to 1. This condition only requires to add probabilities for commuting (orthogonal) projectors, avoiding known objections to other derivations~\cite{laloe}. 
Then all the hypotheses for Gleason's theorem \cite{gleason} are fulfilled (see §~IV), and thus Born's rule applies \cite{CSM_Born2}. By construction orthogonal sets of projectors are connected by complex unitary matrices. Complex numbers are  required to connect continuously the identity matrix to all permutations of modalities: this cannot be done by (real) orthogonal matrices, which split into two subsets with determinants $\pm 1$; 
see \cite{CSM_Born1,CSM_Born2}. $\square$
\vskip 2mm

Here we have considered initial and final modalities, i.e. rank 1 projectors \cite{CSM_Born2}, but more generally Gleason's theorem  provides the probability law for density operators (convex sums of projectors), interpreted as statistical mixtures. This clarifies the  link between Born's rule and the mathematical structure of density operators \cite{Masanes}. 

\section{An overview of Gleason's theorem.}

Gleason's theorem has the reputation of being un-penetrable by physicists, who usually keep away from this frightening monument
(see also Discussion below). Therefore we want to present here a ``physicist's demonstration'', where most mathematical difficulties are deliberately omitted, in order to reveal the big picture. All the (nice) mathematical details can be found in  ``An elementary proof of Gleason's theorem'', by Roger Cooke, Michael Keanes $\&$ William Moran, Math. Proc. Camb. Phil. Soc. 98, 117 (1985) \cite{cooke}, which is more recent and reader-friendly than the original work by Gleason  \cite{gleason}. 

Let us consider a separable Hilbert space $\cal H$ over $\mathbb{R}$ or $\mathbb{C}$, and if dim($\cal H$) = N we denote it  
$\cal C_{\text N}$ (over $\mathbb{C}$) ou $\cal R_{\text N}$ (over $\mathbb{R}$). 
Then we define a  real-valued non-negative function $f$ acting on the unit sphere of $\cal H$, such that for any orthonormal basis 
$\{ x_i \}$ of $\cal H$, one has $\sum_i f(x_i) = 1$. 

The function $f(x_i)$ can be seen as the probability to get the result $x_i$, in a ``state'' defined by $f$.  Note that if $f(x_j) = 1$ for the vector $x_j$, then $f(x_{k \neq j}) = 0$ for all other vectors in the orthonormal basis $\{ x_i \}$: the results $x_i$ are mutually exclusive as we required. The non-obvious hypothesis is why $f(x_i)$ depends only on $x_i$ and $f$, and not on the other vectors $\{ x_{k \neq j} \}$ in the orthonormal basis: this is where the discussion above plays a crucial role, by associating $x_i$ to an extravalence class of modalities.

Here our goal is to sketch a demonstration of Gleason's theorem: { \it If $N \geq 3$, there exists a density operator \footnote{
This means a positive semidefinite Hermitian operator with unit trace. It describes a pure state if it is a rank one projector.} $\rho$ defined on $\cal H$ such that $f(x_i) = \langle x_i | \rho | x_i \rangle$  for all unit vectors $x_i$.}
Then $f$  is said to be ``regular''.  

For simplicity we will assume that the extreme value  $f(x_i) = 1$ is reached,  and  then present the (easier) result that in that case $\rho$ is a projector  $| x \rangle \langle x |$, so $f(x_i) = \langle x_i | x \rangle \langle x | x_i \rangle = |\langle x_i |x \rangle|^2$ : this is the usual Born’s rule for pure states (or for extravalent modalities). 
\vskip 2mm

\noindent  {\bf Step 1:  prove the following ``reduction lemmas''}

\noindent L1 - In $\cal R_{\text N}$,  $f$ is regular  iff  it is the restriction to the unit sphere of a quadratic form (this is clear by writing explicitly $\rho$ as a self-adjoint operator)

\noindent L2 - If $f$ is regular in $\cal R_{\text 3}$, then it is also regular in any 2-dimensional subspace $\cal R_{\text 2}$ of $\cal R_{\text 3}$ (this is clear by restricting the quadratic form from $\cal R_{\text 3}$ to $\cal R_{\text 2}$)

\noindent L3 - If $f$ is regular in any subspace $\cal R_{\text 2}$ of $\cal C_{\text 2}$, then it is regular in $\cal C_{\text 2}$ (not obvious, see \cite{cooke})

\noindent L4 - If $f$ is regular in any subspace $\cal C_{\text 2}$ of $\cal C_{\text N}$, then it is regular in $\cal C_{\text N}$ (not obvious, see \cite{cooke})

\noindent {\bf Crucial lemma:}  If $f$ is regular in $\cal R_{\text 3}$, then it is regular in $\cal C_{\text N}$  (use L2, then L3, then L4). 

Therefore it is enough to show that $f$ is regular in $\cal R_{\text 3}$. This explains why the theorem requires $\text{N} \geq 3 $: in fact, $f$ is regular in any $\cal C_{\text 2}$ considered as a subspace of $\cal C_{\text N}$, but not in $\cal C_{\text 2}$ considered alone. Said otherwise, it is well known, e.g. from Clauser \cite{clauser}, that one can build a ``classical model of a (unique) qubit''. However this classical model fails if this qubit is one among several qubits, which is fine as far as  QM is concerned. 
\vskip 2mm

\noindent  {\bf Step 2: prove that $f$ is regular in $\cal R_{\text 3}$}

\noindent Now one looks for a probability function  $f(u)$, where $u$ is a normalized vector in $\cal R_{\text 3}$, so that 
$0 \leq f(u) \leq 1$ and  $f(u)+f(v)+f(w)=1$ for any orthonormal basis $\{u,v,w\}$  of $\cal R_{\text 3}$.  One does not assume that $f$ is continous, but here we assume that the extreme values  0 and 1 are reached (this is only for simplification, and the general case is treated in the full theorem \cite{gleason,cooke}).

Given a normalized vector $p$ and an orthonormal basis $\{u,v,w\}$, the quantities $\cos^2(u,p)$, $\cos^2(v,p)$, $\cos^2(w,p)$ are the squares of the components of $p$ in the basis, so they sum to 1. Therefore $\cos^2(u,p)$ for a fixed $p$ is an acceptable function $f(u)$, and actually it is the good one. But why is it the only such function? We will split the answer in two parts. 

\begin{figure}[h]
\includegraphics[width = 3.5cm]{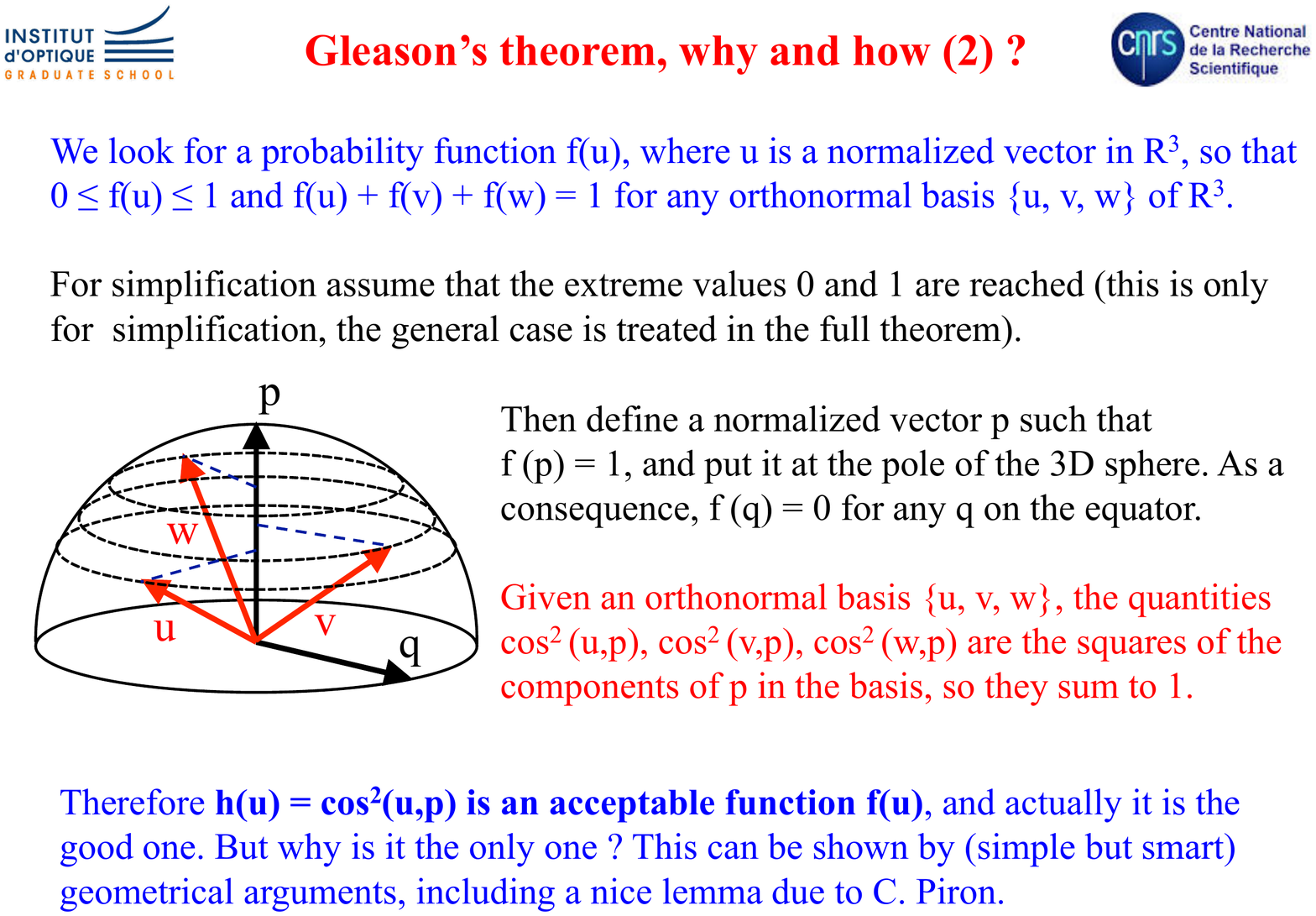}
\caption{We look for a probability function $f(u)$, where $u$ is a normalized vector in $R^3$, so that $0 \leq f(u) \leq 1$ and $f(u) + f(v) + f(w) = 1$ for any orthonormal basis $\{u, v, w \}$ of $R^3$. To be simple (see text) we assume that the extreme values 0 and 1 are reached, we define a normalized vector $p$ such that $f(p) = 1$, and put it at the pole of the 3D sphere. As a consequence, $f(q) = 0$ for any $q$ on the equator. Given an orthonormal basis $\{u, v, w \}$, the quantities $\cos^2(u,p)$, $\cos^2(v,p)$, $\cos^2(w,p)$ are the squares of the components of $p$ in this basis, so they sum to 1. Therefore $h(u) = \cos^2(u,p)$ is an acceptable function $f(u)$, and Gleason's theorem shows that it is the only one. }
\end{figure}

\noindent  {\it Why is there no $\phi$ ? }
\vskip 1mm

In $\cal R_{\text 3}$ a normalized vector $u$ is defined by two polar angles $\theta$ and $\phi$, and in  $\cos^2(u,p)$ there is only one angle, why? To be specific let us choose $p$ as the vector such that $f(p) = 1$ (since this value is reached), and position it at the pole of the unit sphere in 3 dimensions (see Fig.~2). As a consequence, $f(q) = 0$ for all vectors on the equator, and for any vector one can define $h(u) =\cos^2(u,p)$, which depends on the ``latitude" of $u$ (the polar angle $\theta$), but not on its ``longitude" (the azimuthal angle $\phi$). 

One can then use two lemmas  
to show that for any two vectors $u$, $v$ in the northern hemisphere such that $h(u) > h(v)$, one has $f(u) \geq f(v)$; this is done in Annex 1. Then define the smallest and largest values of the possible values of $f(u)$ for a given latitude:
\begin{eqnarray}
m(x) = \text{Inf} \{ f(u) \text{ such that } h(u) = x \}  \nonumber \\ 
M(x) = \text{Sup} \{ f(u) \text{ such that } h(u) = x \}.\nonumber
\end{eqnarray}
One has $M(1) = m(1) = 1$ at the pole, $M(0) = m(0) = 0 $ at the equator, 
and if $x < x'$ then $M(x) \leq m(x')$ due to the above lemmas. In addition one has obviously $m(x) \leq M(x)$ and $m(x') \leq M(x')$, 
so if $x \rightarrow x'$ one gets a contradiction ($M$  less than $m$), unless
$m(x) = M(x)$, i.e. $f(u)$ depends only on the latitude. $\square$
\vskip 2mm

\noindent  {\it Why only $\cos^2(u,p)$ ? }
\vskip 1mm

Given that $\cos^2(u,p)$ is an acceptable $f(u)$, one may think that any other function $f(u) = g( \cos^2(u,p) )$ should be acceptable also. To show this not the case, one uses a 
\vskip 1mm

\noindent {\bf Magical lemma:}  Consider a function $g$ over  [0, 1], verifying the hypotheses  (i) g(0)=0, (ii)  $a < b \Rightarrow g(a) < g(b)$,  (iii) $a+b+c=1 \Rightarrow g(a)+g(b)+g(c)=1$. Then $g(a) = a $ for any $a$ within [0,1].

The proof (subtle but not difficult) is given in Annex 2.  It  is easily seen that $g( \cos^2(u,p) )$ fulfills the hypothesis of the lemma for any orthonormal basis  $\{u,v,w\}$, with $a =  \cos^2(u,p)$ etc. So from the magical lemma one gets $g(\cos^2(u,p)) =  \cos^2(u,p)$, and the additional function $g$ is useless.
$\square$
\vskip 2mm

\noindent  {\bf Step 3: conclude that $f$ is regular in $\cal C_{\text N}$}

\noindent Therefore $f$ is regular in $\cal R_{\text 3}$, and also in $\cal C_{\text N}$ from the reduction lemmas. The demonstration can be reconsidered in the more general case where the value $f(p) = 1$ is not reached,  and 
one finds\footnote{In the general case in $\cal R_{\text 3}$,  the maximum (resp. minimum) value of $f$ is $0 \leq M  \leq 1$ (resp. $0  \leq m  \leq 1$), and one shows \cite{cooke} that there exist a basis  $\{ p, q, r \}$ such that $f(u) = M \cos^2(u, p) + m \cos^2(u, q) + (1-M-m) \cos^2(u, r)$  with $M+m  \leq 1$.} that $\rho$ is no more a projector, but a density matrix associated with a statistical mixture. $\square$

\section{Discussion.}

An essential feature of the contextual quantization postulate, i.e. the fixed value $N$ of the maximum number of mutually exclusive modality, turns out to be the dimension of the Hilbert space. In the spirit of \cite{lipton} and as shown is \cite{CSM_Born1,CSM_Born2},  this provides one more heuristic reason for using projectors. Then the projective structure of the probability law warrants that, despite the availability of an infinite number of incompatible modalities,  $N$ cannot be ``bypassed'' by getting more details on any of them. 

This would not be the case in the usual probability theory, based on partitions : making a partition of all modalities  in $N$ sub-ensembles for each given context would not prevent sub-partitions, that would correspond to additional details or ``hidden variables'', that are forbidden by our basic postulate. This corresponds mathematically to Bell's or  Kochen-Specker's  theorems, and all their variants, which basically show the inadequacy of partition-based probabilities.  This problem obviously vanishes when projectors are used, and then from Gleason's theorem no other choice is left than Born's rule. 
It is worth emphasizing also that Bell's or  Kochen-Specker's  theorems consider discrete sets of contexts, whereas Gleason's theorem is based upon the interplay between the continuum of  contexts, and the quantized number of modalities accessible  in a given context. This feature also fits perfectly with the CSM ideas. 

We note that some recent derivations of Born's law \cite{Zurek,Cabello,Masanes} dismiss Gleason's theorem, on the basis that its hypotheses are either too strong (extracontextuality) or unjustified (projective probabilities). More precisely, refs. \cite{Zurek,Cabello,Masanes} argue for the non-relevance of Gleason's theorem to QM, in opposition to the CSM view.  Quoting \cite{Masanes}: {\it ``As mentioned in the introduction, Gleason's theorem and many other derivations of the Born rule assume the structure of quantum measurements. That is, the correspondence between measurements and orthonormal bases $\{ \varphi_i \}$, or more generally, positive-operator valued measures. But in addition to this, they assume that the probability of an outcome $\varphi_i$ does not depend on the measurement (basis) it belongs to.'' }  

In~\cite{Masanes} this additional assumption (which is physically true) is called ``non-contextuality'', that is clearly misleading, clashing with the terminology used in the Kochen-Specker theorem. As written above, a better name is ``non-contextual assignment of probabilities'', and the best name is just extracontextuality, that has deep physical roots. This is made clear by associating projectors to extravalence classes, clearly distinguishing the physical  result (the modality) and the mathematical construction (the projector). To answer the remark about ``assuming the structure of quantum measurements'', we do posit the projective structure of quantum probabilities \cite{CSM_Born2}, not as a deduction but as a duly justified  inference \cite{lipton}. In the CSM approach the mathematical formalism works because physics tells the rules, and not the opposite.

Therefore in our approach Gleason's  hypotheses have a deep physical content, linking contextual quantization and extracontextuality of modalities. Since these features are required from empirical evidence,  the QM formalism provides a good answer to a well-posed question.

\section{Algebraic scheme for quantum measurements. }

A consequence of our approach is that  usual textbook quantum mechanics, which is limited to type-I operator algebra as introduced initially by Murray and Von Neumann \cite{MVN}, is not universal because it does not include the context. This issue was already discussed by Von Neumann  \cite{JvN1939}, and again later in the framework of algebraic quantum theory \cite{Landsman}. 
Nevertheless, as discussed in these articles, it is possible to get a full picture by including the context in the formalism, taking into account that its number of degrees of freedom is unbounded, which makes its algebra of operators non-type I \cite{JvN1939,Landsman}. 

\begin{figure}[h]
\begin{center}
\includegraphics[width = 8.5cm]{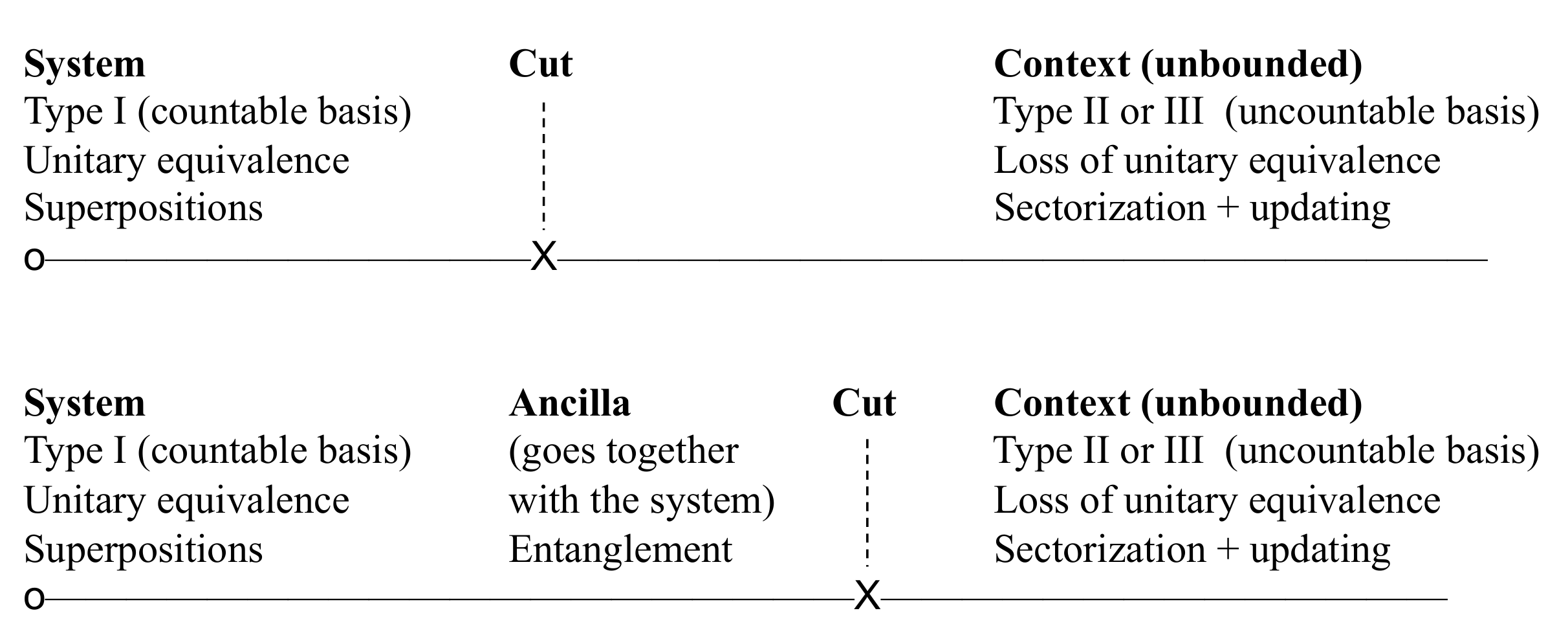}
\end{center}
\caption{Generic scheme including the system, a possible ancilla, and the context. The number of degrees of freedom in the context is unbounded, which makes its algebra non type-I. The cut separates a type-I system algebra, where usual QM applies, from the type-II or III context algebra where there is no more unitary equivalence of representations.  }
\label{mesgen}
\end{figure}

Fig. \ref{mesgen} displays such a generic scheme, including the system (plus ancillas) and context, separated by a (movable) cut.  The full scheme is then universal, but the mathematical description including type II or III algebra does not allow arbitrary quantum superpositions at the context level -- in agreement with empirical evidence. 
Then a quantum measurement proceeds as follows:
\begin{itemize} 

\item Before the measurement the modality is associated with  the following (density) operator in context $C_1$:    
$$| \psi_i \rangle \langle \psi_i |  \otimes  \rho^{(C_1)}_i $$
Specifying the modality requires to give {\bf both}  $| \psi_i \rangle \langle \psi_i |$ and $\rho^{(C_1)}_i$ 
 because the projector $| \psi_i \rangle \langle \psi_i |$ specifies only an extravalence class of modalities. 
 
\item After the measurement carried out in context $C_2$, but before reading out the result, the sectorized state (statistical mixture) is
$$  \sum_j \;  p_j  \;  | \phi_j \rangle \langle \phi_j |  \otimes  \rho^{(C_2)}_j$$
This form is completely generic from a mathematical point of view because the context is unbounded, and it can be justified in several possible ways: sectorization in the non-type-I algebra, loss of off-diagonal elements of the reduced density matrix, flow of information to the environment, loss of interference, loss of the ability to create entanglement in a projective measurement... They all lead to the same results, as discussed e.g. in  \cite{coles}. 

\item After reading out the measurement result $k$ in context $C_2$, the new modality can be updated and it
 is associated with  the  operator:     
$$ | \phi_k \rangle \langle \phi_k |  \otimes  \rho^{(C_2)}_k$$
This defines a new pre-measurement modality, and $ | \phi_k \rangle \langle \phi_k |$ may evolve unitarily until the next measurement is performed. 
\end{itemize}

Summarizing, the non-unitary step in the measurement is due to the fact that the whole unbounded context is involved in a transient way; this is not an additional ingredient, but a required part of the full (non type-I) formalism. Looking at $ | \psi \rangle$ as the  ``state of the system", as done usually,  is misleading because the vector (or projector)  is associated with an extravalence class of modalities. 
The basic CSM tenet, that the modality belongs to both the system and the context, appears explicitly here under a mathematical form.  

As a conclusion, usual type I QM provides a description of (idealized) isolated quantum systems. A state vector or projector is ``incomplete" because it is not associated with an actual modality, but with an extravalence class of modalities, belonging to different contexts. From a physical point of view, the modality belongs jointly to a quantum system, and to a specified context. From a mathematical point of view, the behavior of modalities can be studied using type-I QM, where Born's rule applies as a consequence of Gleason's theorem. On the other hand, the description of (unbounded) contexts requires a non type-I formalism. Overall, these combined tools provide a consistent picture of quantum measurements within a unified quantum framework. 
\\

\noindent {\bf Annex 1 : Proof of the geometrical lemmas.} 
\vskip 1mm

Here we  show that for two vectors $u$, $v$ in the northern hemisphere with $h(u) > h(v)$, one has $f(u) \geq f(v)$. 

For this purpose we define $D_u$, the great circle going through $u$ and cutting the equator at two points corresponding to vectors orthogonal to $u$. By convention $D_u$ is called the ``descent through $u$", and $u$ is obviously the ``northern vector" in $D_u$. Then one prooves the two lemmas: 
\vskip 2mm

\noindent {\it Basic lemma :}  One has  $f(u) \geq f(u')$ for any $u'$ in $D_u$. 
\vskip 1mm

\noindent {\it Proof : } Consider a vector $u$, and another vector $u'$ within $D_u$. Let $v$  (resp. $v'$) be a vector in $D_u$ orthogonal to 
$u$ (resp. $u'$). Adding a vector $w$ perpendicular to the $D_u$ plane, $\{u, v, w \}$ and $\{u', v', w \}$ are two orthonormal basis. By definition of $f$ one has 
$f(u) + f(v) + f(w)  = f(u') + f(v') + f(w)$ therefore $ f(u) = f(u') + f(v')$ since $f(v) = 0$ because $v$ is on the equator. 
Since $ f(v') \geq 0$ one has $f(u) \geq f(u')$. $\square$
\vskip 2mm

\noindent {\it Piron's lemma :} Consider  $u$, $v$ such that $ h(u) > h(v)$. Then there is a series of $N$ vectors $w_n$ such that $w_0 = u$, $w_N = v$, and each $w_n$ is within $D_{w_{n-1}}$, i.e. in the descent through the previous vector of the series. 
\vskip 1mm

\noindent {\it Proof :} It relies on a smart geometrical construction due to Piron \cite{piron}. It is convenient to project the northern hemisphere on a plane tangent  at the pole $p$, using a projection from the center of the sphere. The different latitudes are then concentric circles centered on $p$, and the equator is projected at infinity. The descent through $u$ is a straight line, tangent at $u$ to the circle corresponding to the latitude of $u$. Then there are two cases :

- If $u$ and $v$ have the same longitude, one takes $u=w_0$, $v = w_2$, and there exists $w_1$ with a latitude between those of $w_0$ and $w_2$, located on $ D_u=D_{w_0}$, and such that $w_2$ is on $D_{w_1}$  (this is clear by looking at the previous projection, $u$ and $v$ are on the same line coming from $p$).

- If $u$ and $v$ have different  longitudes, one can take $u=w_0$, $v = w_N$, and build the other vectors $w_n$ by progressively rotating between the two circles associated to the two latitudes. When these latitudes get closer, $N$ becomes larger, and it tends to infinity for two different longitudes with almost the same latitude (again this is clear from a drawing). This proves the lemma. 
\vskip 2mm

Therefore the basic lemma relates $u$ and $u'$ within the descent through $u$, and Piron's lemma relates $u$ and $v$ from a succession of descents through the vectors in the series $w_n$. As a conclusion, one deduces from the two lemmas that for  $u$, $v$ in the northern hemisphere with $h(u) > h(v)$, one has $f(u) \geq f(v)$. $\square$
\vskip 5mm

{\bf Annex 2 : Proof of the magical lemma.} 
\vskip 1mm

For mathematical reasons related to continuity \cite{darboux},
it is assumed that $g$ is defined on [0,1] except an at most countable set $K$ of points, and that the hypotheses (i)~g(0)=0, (ii)~$a < b \implies g(a) < g(b)$,  (iii)~$a+b+c=1$ $\implies g(a)+g(b)+g(c)=1$ are valid under these same conditions.  One has then $g(1) = 1$,
since  $g(0) = 0$ and $g(a)+g(b)+g(c)=1$. 
Considering the rational numbers $r$  and $s$, and $a_0$ outside $K$, one gets:
\begin{eqnarray}
g(r a_0) + g(s a_0) + g(1 - r a_0 - s a_0) = 1 \nonumber \\ 
g(0) + g(r a_0 + s a_0) + g(1- r a_0 - s a_0) = 1 \nonumber \\ 
\implies    g(r a_0) + g(s a_0) = g(r a_0 + s a_0) = g((r + s) a_0) \nonumber \\ 
\implies  g(r a_0) = r g(a_0)   \nonumber
\end{eqnarray}
Taking the limit $r \rightarrow 1/a_0$ one gets $g(1) = g(a_0) / a_0 = 1$
and thus $g(a_0) = a_0$  and   $g(r a_0) = r a_0$.    $\square$

\begin{acknowledgements}
\vskip -3mm 
The authors thank  Franck Lalo\"e and Roger Balian for many useful discussions, 
and Nayla Farouki for continuous support.
\end{acknowledgements}

%
%



\begin{thebibliography}{}
\bibitem{darboux} A real-valued monotone function on [a,b] is continuous, except in an at most countable set of discontinuities (Froda-Darboux theorem).
\end{thebibliography}
\end{document}